\documentclass[prb,preprint]{revtex4}

\usepackage{amsfonts,amssymb,eucal}
\usepackage{dcolumn}
\usepackage[ansinew]{inputenc}
\usepackage[centertags]{amsmath}
\usepackage{graphicx}
\usepackage{times}
\usepackage{color}

\begin{document}
\bibliographystyle{aip}
\newcommand{\bra}[1]{\ensuremath{\langle #1 \vert}}
\newcommand{\ket}[1]{\ensuremath{\vert #1  \rangle}}
\newcommand{\braket}[2]{\ensuremath{\langle  #1\vert #2  \rangle}}
\newcommand{\ketbra}[1]{\ensuremath{\vert{#1}\rangle\langle{#1}\vert}}
\newcommand{\be}{\begin{equation}}
\newcommand{\ee}{\end{equation}}
\newcommand{\bea}{\begin{eqnarray}}
\newcommand{\eea}{\end{eqnarray}}
\newcommand{\vp}{\varphi}
\newcommand{\h}{\hat}
\newcommand{\ve}{\varepsilon}
\newcommand{\vrr}{\vec r}
\newcommand{\pr}{\;\prime}
\newcommand{\noi}{\noindent}
\date{}

\markright{}
\title{A Connection between Special Theory of Relativity and Quantum Theory}
\author{I. Mayer}
\affiliation{Research Centre for Natural
Sciences\footnote{Formerly called Chemical Research Center},
Hungarian Academy of Sciences,
H-1525 Budapest, P.O.Box 17, Hungary}
\email{mayer@chemres.hu}
%
%
%
%
%

\date{\today}

\begin{abstract}

The special theory of relativity does not predict the existence of photons
(quanta of electromagnetic radiation). However, it is demonstrated here that
it follows from the special theory of relativity that
{\it if photons do exist}---and we know that 
they do---then their energy must be proportional to 
their frequency. This means that the Planck-Einstein formula \mbox{$E=h\nu$}
follows just from some results  of the special theory of relativity
{\it and\/} the assumption of the particle--wave duality.


\end{abstract}

\maketitle
%

%

%
%


\subsection*{I. Introduction}

The two revolutionary physical theories of the early XX-th century, the
special theory of relativity (STR) and quantum theory, were developed
in parallel, and until now they are often considered somewhat independently of
each other, except in the highly sophisticated theories like quantum 
electrodynamics. In particular, the Planck-Einstein formula $E=h\nu$ 
for the quanta of electromagnetic waves (photons) is usually considered as a
postulate deduced from Einstein's interpretation of the black-body radiation 
and photoelectric effect, \cite{Einstein-ph} without looking for any 
direct connection with the results of STR. 

In fact, there are important relationships in Maxwellian 
electrodynamics---an integer part of STR---which can be considered 
as anticipating the existence 
of photons. For instance: 
``the relation between energy $W$ and momentum $W/c$ for the 
electromagnetic wave is the same as for a particle moving with the velocity
of light".\cite{LL} That, however, does not tell anything about the 
relationship between the energy of that ``particle'' and the frequency of
the electromagnetic wave. The wave equation
\be
\Delta f - \frac1{c^2}\frac{\partial^2 f}{\partial t^2} =0 \quad,
\ee
is satisfied for every twice differentiable function of the form
$f({\bf kr}-\omega t)$, so it cannot give information about the possible
``wave packets" either.

Nonetheless, we are going to demonstrate here that the relationship 
\mbox{$E=h\nu$}
follows directly from the results of STR, if one augments it with the 
assumption of the particle--wave duality.

\subsection*{II. Transformation of the electromagnetic energy}

We start with a very simple derivation of the formula connecting the 
energies of the electromagnetic wave in two inertial reference systems.

Let us consider two inertial systems of reference $K$ and $K^\prime$,
and assume that system $K^\prime$ moves with the velocity $V$ in the positive
direction of the axis $x$ with respect to the system $K$. 
We assume that the axes of the two system
are parallel and that 
the electromagnetic wave also propagates along the axes $x,\ x^\prime$.


The spatial energy density $W$ of the electromagnetic filed expressed in the
symmetric Gaussian system is, as known \cite{LL}
\be
W=\frac{E^2 + H^2}{8\pi} \quad,
\ee
where $E$ and $H$ are the intensities of the electric and magnetic fields,
respectively. (For a plane wave one has $|\vec E|=|\vec H|$.) The spatial
energy density transforms\footnote{One can 
easily obtain the same result by considering
the Lorentz-transformation of the components of the electromagnetic field,
by selecting the relative signs of the components $E_y,\ H_z$ and $E_z,\ H_y$
in such a manner as to provide the vector of the energy flux
$\vec S = (c/4\pi)[\vec E \vec H]$ 
to be directed along the positive $x$ axis.}
under Lorentz-transformation as\cite{LL} 
\be
W = W^\prime\frac{\left(1+\frac{V}{c}\right)^2}{1-\frac{V^2}{c^2}}\quad,
\ee
where we have substituted $\cos\alpha=1$ for the angle $\alpha$ 
between the direction of the wave propagation and the axis $x$, as they were
assumed parallel for the sake of simplicity. 

The four-dimensional wave vector 
${\bf k}=\{\frac{\omega}c,k_x,k_y,k_z\}$
transforms as other 4-vectors: \cite{LL}
\be
k_0 = \frac{k_0^{\,\prime} +\frac{V}{c}k_x^{\,\prime}}{\sqrt{1-\frac{V^2}{c^2}}}
\quad .
\ee
Substituting here $k_0=\omega/c=2\pi\nu/c=2\pi/\lambda$ and 
$k_x=2\pi/\lambda$,
one gets\footnote{$k_y=k_z=0$ in the given case.}
\be
\frac{\nu}{\nu^\prime}=\frac{\lambda^\prime}{\lambda}=\frac{1+\frac{V}{c}}
{\sqrt{1-\frac{V^2}{c^2}}}\quad .
\ee

The energy ${\cal E}$ and ${\cal E}^{\,\prime}$ of a  pulse of  
monochromatic radiation, consisting of a finite number
of periods 
can be obtained in systems $K$ and $K^\prime$, respectively, by integrating 
$W$ and $W^\prime$ over a given
number of waves (phases $2\pi$), depending on the actual form of the pulse:
\be
{\cal E} = \int W dv \ ; \quad
\quad {\cal E}^{\,\prime} = \int W^\prime dv^{\,\prime}  \quad.
\ee
When comparing ${\cal E}$ and ${\cal E}^{\,\prime}$, one should take into 
account that they correspond to {\it the same\/} pulse considered
in different reference systems, so the integration is under the same number
of waves (wave lengths, full phases $2\pi$). At the same time, $W$ gives
the density of the electromagnetic energy in the three-dimensional space.
The longer is the wave length, the larger is the integration 
volume. (The elements of the length in the orthogonal directions do 
not change.) 
Therefore,
one should consider that the elements of the volume $dv$ and $dv^{\,\prime}$ 
relate to each other as the wave lengths $\lambda$ and $\lambda^{\;\!\prime}$.
As the consequence, the energy of the pulse in the two inertial
systems will relate as
\be
\label{arany}
\frac{{\cal E}}{{\cal E}^{\,\prime}}=
\frac{W}{W^\prime}\frac{\lambda}{\lambda^{\;\!\prime}}=
\frac{\nu}{\nu^{\,\prime}} \quad.
\ee

\subsection*{III. The Planck-Einstein formula}

The result  Eq.~(\ref{arany}) indicates that in each 
inertial system the energy of 
a pulse of electromagnetic radiation formed of monochromatic waves 
is proportional to the frequency of the wave
in that reference system. Now we assume, following Einstein's proposition,
the particle--wave duality, and consider the pulse of radiation as consisting
of a given number ${\cal N}$ of quanta (photons) of equal energy ($E$ and
$E^{\,\prime}$ in the two systems, respectively).
Supposing also that {\it the number of the quanta ${\cal N}$  is a relativistic 
invariant}, we can write Eq. (\ref{arany}) as
\be
\label{arany1}
\frac{{\cal E}}{{\cal E}^{\,\prime}}=
\frac{{\cal N}E}{{\cal N}E^{\,\prime}}=
\frac{E}{E^{\,\prime}}=
\frac{\nu}{\nu^{\,\prime}}\quad.
\ee
The assumption of the wave--particle duality is really meaningful and
permits to consider the photons as a  well-defined sort of
particles only if
the energy of the photon depends only on its frequency
but not on the condition of its emission, polarization {\it etc.}. 
That means that it can be characterized by
a {\it universal function\/} of the frequency (and only of the frequency)
$E=E(\nu)$. Then one can write $E=E(\nu)$ and
$E^{\,\prime}=E(\nu^{\,\prime})$ in Eq.~(\ref{arany1}), so it reduces to
\be
\label{arany2}
\frac{E\;\!(\nu\;\!)}{E(\nu^{\,\prime})}=
\frac{\nu}{\nu^{\,\prime}}\quad.
\ee
This equality should be fulfilled for any inertial system, so for any pairs
of frequencies
$\nu$, $\nu^{\,\prime}$, which is only possible if the energy of the photon $E$
is proportional to its frequency $\nu$.
Thus  we arrived at 
the Planck-Einstein formula 
\be
\label{PE}
E=h\nu \quad,
\ee
as a consequence of special theory of relativity. (Of course, the actual value
of the Planck's constant $h$ cannot be determined from STR alone.) 
One can also argue that the relationship Eq.~(\ref{arany}) is simply 
necessary to avoid a conceptual conflict between STR and the
notion of the light consisting of quanta of the energy $h\nu$.


Obviously, the conclusion that the Planck-Einstein formula follows
directly from STR {\it and\/} the particle-wave duality
could be obtained, say, some 70-80 years ago.
But apparently it was not.
At the same time it 
may have significant pedagogic power as well as illustrates that our
basic physical theories are in harmony with each other.

Of course, the simple 
result of Eq.~(\ref{arany}) is by far 
not new. It is included, among others, in the textbook of Novozhilov and 
Yappa. \cite{Nov-Japp} These authors also discuss that this result is in 
agreement with the Planck-Einstein formula, as it is postulated by the 
quantum theory, but do not make the conclusion 
that this result together with the assumption of the particle-wave duality 
{\it necessarily\/} leads to that formula, 
thus there is no need of postulating it 
empirically. We think this to be of significant conceptual importance.


%

\subsection*{IV. Conclusions}
It is demonstrated here 
that the Planck-Einstein formula \mbox{$E=h\nu$} follows 
just from some standard results of the special theory of relativity
and the assumption of the particle--wave duality: the energy of a monochromatic
pulse of electromagnetic radiation transforms proportionally to its frequency.
Consequently, if the pulse consists of a given number of 
quanta of equal energy (photons) then
the energy of each quantum must be proportional to the frequency.

\subsection*{V. Addenda}

After the first version of this manuscript has been uploaded to arXive
preprint server, 
Dr. John H. Field (University of Geneva) has informed me about his paper 
\cite{Field} with a different derivation of Eq.s (\ref{arany}) and (\ref{PE}). 
He called the attention to the fact that the result of Eq. (\ref{arany})
was already 
obtained by Einstein in his first paper about STR \cite{Einstein-STR}.
Einstein's derivation was based on considering the energy of the ``light 
complex'' 
included in a sphere that in other reference system becomes an ellipsoid.
Einstein concluded:

``It is remarkable that the energy and frequency of a light complex vary 
with the state of motion of the observer in accordance with the same law."

This remark might be a hint to the relationship between the energy and 
frequency of the quanta of light that were introduced \cite{Einstein-ph} 
by himself several months {\it before} the STR paper \cite{Einstein-STR}, 
but that connection was not done explicitly, and no reference was made 
at all to the previous paper.
 
It is also to be mentioned that several people called my attention  
that 
the result of Eq.~(\ref{arany})  immediately follows also from the fact that
the energy ${\cal E}$ and the frequency $\nu$ are both the zero-th components 
of parallel
light-like vectors pertinent to the monochromatic pulse.

\subsection*{Acknowledgment}
The author thanks Dr.~Gy\"orgy Lendvay for helpful discussions and 
Prof. Tam\'as Geszti for a useful exchange of e-mails.

%

%
%
%
%
%
%
\end{document}